\documentclass[aps,twocolumn,floats,nofootinbib]{revtex4}
\usepackage{amssymb,color}
\usepackage{amsmath}

\newcommand{\beq}{\begin{equation}}
\newcommand{\eeq}{\end{equation}}
\newcommand{\beqn}{\begin{eqnarray}}
\newcommand{\eeqn}{\end{eqnarray}}

\begin{document}

\title{Information processing in living systems}

\author{Ga\v{s}per Tka\v{c}ik$^{1}$\footnote{gtkacik@ist.ac.at}  }

\author{William Bialek$^{2}$\footnote{wbialek@princeton.edu}  }

\affiliation{\mbox{${}^1$Institute of Science and Technology Austria, Am Campus 1, A-3400 Klosterneuburg, Austria} \\ ${}^{2}$Joseph Henry Laboratories of Physics and Lewis--Sigler Institute for Integrative Genomics,\\ Princeton University, NJ 08544 Princeton, USA}

\date{\today}

\begin{abstract}
Life depends as much on the flow of information as on the flow of energy.  Here we review the many efforts to make this intuition precise.  Starting with the building blocks of information theory, we explore examples where it has been possible to measure, directly, the flow of information in biological networks, or more generally where information theoretic ideas have been used to guide the analysis of experiments. Systems of interest range from single molecules (the sequence diversity in families of proteins) to groups of organisms (the distribution of velocities in flocks of birds), and all scales in between.  Many of these analyses are motivated by the idea that biological systems may have evolved to optimize the gathering and representation of information, and we review the experimental evidence for this optimization, again across a wide range of scales.
\end{abstract}

\maketitle
\section{Introduction}
Cells and organisms sense, compute, and make decisions: to proliferate, to find food, to protect themselves against predators and unfavorable environmental changes, to act in unison with their neighbors within a collective, and---broadly speaking---to invest their limited resources to their maximal benefit. These processes span a wide range of temporal and spatial scales, and reflect dynamics in phase spaces of widely varying intrinsic dimensionality. In bacteria, for example, the  chemotactic signaling network comprises just a handful of chemical species, responds to changes in the local nutrient concentration on  the one second timescale, and modulates bacterium's swimming by controlling the direction of flagellar rotation.  At the other extreme of complexity, millions of neurons in the brain of a pianist are engaged in processing roughly one megabit per second of visual, auditory, and tactile information, combining these data with the score retrieved from memory, and computing the fine motor program controlling the pianist's fingers. Other processes take place over much longer timescales: DNA sequences in a population of organisms are subject to the evolutionary forces of mutation, selection, and random drift, and changes in the environment slowly shape the distribution of genomic sequences. 
In all these cases, intuition tells us that information is ``flowing'' from the outside environment to a representation internal to the organism or a population, and that this information flow is essential for life.  Can this intuition be formalized? If so, can it be  connected, quantitatively, to experimental data? Most ambitiously, can a focus on information flow help us to unify our description  of the complex and diverse phenomena that we see in the living world?

The  purpose of this review is to explain how the intuition about ``biological information flows'' can be formalized in the language of information theory, and what the resulting research program  teaches us about the physics of living systems. We introduce the basic concepts of information theory (Section~\ref{sec2}), show how this mathematical structure leads to new views of experimental data (Section~\ref{sec3}),  and explore the possibility that some of life's mechanisms can be understood as solutions to the problem of optimizing information flow subject to physical constraints  (Section~\ref{sec5}).   Throughout, we follow the physics tradition of emphasizing the unity of theoretical ideas at the expense of some of the biological complexity. 

Before we begin, it seems worth discussing how ideas about information flow relate to the more familiar theoretical structures in physics. Intuitions about information flow are not confined to biological systems.    When we look a ferromagnet, for example, we know that all the details of the electronic configurations on one side of the sample are not relevant if we try to predict those on the other side of the sample; what does  carry information across long distances is the order parameter, in this case the magnetization.  But precisely because we can identify the order parameter, we don't need to speak abstractly about information.  Rather  than trying to turn ``carry information'' into a precise mathematical statement, we just compute correlation functions of the order parameter evaluated at distant points. 

For most of the biological systems that we will be discussing, finding something like an order parameter remains a distant hope.  Symmetries are absent, and even locality is not much of a guide to understanding, for example,  a network of neurons in which each cell may be connected to many thousands of neighbors.  Under these conditions, some more general approaches to formalizing our intuitions about information flow would be useful.  Interestingly, as we write this, information theoretic ideas are becoming more important in the study of topological phases in correlated electron systems, where no local order parameter is possible.
 
\section{Building blocks of information theory} 
\label{sec2}

Here we give a brief exposition of information theory, focusing on topics that are most relevant for our subsequent discussion; see also Ref \cite{bialek_12}.   The foundational papers by Shannon are wonderfully readable \cite{shannon_48}, there is an excellent textbook account aimed at the core of the subject \cite{cover+thomas_91}, and a more recent text that emphasized connections between ideas information theory and statistical physics \cite{mezard+montanari_09}.  Ideas which have been most widely used in thinking about biological systems are not necessarily those that have been most useful in the (enormously successful) application of information theory to communication technology, so our perspective here is somewhat different from that in Ref \cite{cover+thomas_91}.

When we ask a question, and hear the answer, we gain information.  In 1948 Shannon  asked whether we could attach a number to the amount of information we gain \cite{shannon_48}.  He imagined that, upon asking the question, we could envision a set of possible answers (${\rm n} = 1,\, 2,\, \cdots ,\, N$) and that we could assign probabilities ${\mathbf p} \equiv \{p_1, \, p_2,\, \cdots ,\, p_N\}$ to each of these answers. Any acceptable measure of information must obey some constraints:  if all $N$ possible answers are equally likely, then the information must grow with $N$; if the question can be decomposed into independent parts, then the information gained on hearing the answer to each part should add to the total; if we can decompose the full question into a tree of choices (as we do when playing twenty questions), then the total information should be the weighted sum along the paths through the tree.  Remarkably, these postulates allow only one possible measure of information, the entropy of the distribution of answers,\footnote{Shannon's theorem makes precise the vague statement made in many statistical mechanics classes, that entropy is related to our lack of information about the microscopic state of a system.  The more positive version is that the entropy is the information we would gain upon learning the full microscopic state.} $S[{\mathbf p}]= -k\sum_{\rm n} p_{\rm n}\log p_{\rm n}$. 

A seemingly very different question concerns the amount of space required to write down, or represent the answer to our question.  While each possible answer might require a different amount of space, if this is a question we can ask many times (what will the temperature be at noon tomorrow?) then it makes sense to ask about the minimum amount of space required, per answer, in the limit of a long sequence of question/answer pairs.   The answer is again the entropy.  There is some ambiguity about units---the number of characters we need to write the answer depends on our choice of alphabet, and the entropy itself has an arbitrary constant $k$---but if we choose, for example, a binary alphabet, we can set $k=1$ and $\log \equiv \log_2$, then all of the ambiguity is resolved.  When entropy or information is measured in these units, they are called `bits.'

Getting the precise answer to a question is a rare thing.  More common is that we are interested in $x$ but can only measure $y$.  Before a measurement we know only that $x$ is drawn from the distribution $P_X(x)$; after the measurement we know that it is drawn from the conditional distribution $P(x|y)$.  The same arguments that establish the uniqueness of the entropy now show that the only acceptable measure of the information about $x$ gained by observing $y$ is the reduction in entropy, 
\begin{eqnarray}
I (y\rightarrow x) &=& -\sum_x P_X(x) \log_2 P_X(x) \nonumber\\
&&\,\,\,\,\, - \left[  -\sum_x P(x|y) \log_2 P(x|y) \right].
\label{IdiffS1}
\end{eqnarray}
If we ask not about the information gained from observing a particular $y$, but the average over all the $y$'s we can observe, this average information is actually symmetric in $x$ and $y$, and is usually called the ``mutual information.''  It can be written in a manifestly symmetric form, and unlike the entropy itself it is well defined even when $x$ and $y$ are continuous variables,
\begin{equation}
I(x;y) = \iint dx\;dy\; P(x,y) \log_2\left[\frac{P(x,y)}{P_X(x)P_Y(y)}\right] , \label{info1}
\end{equation}
where $P(x,y)$ is the joint distribution of $x$ and $y$.  This can be extended to the case where $x$ and $y$ are multidimensional, or even functions, in which case the sums over states become functional integrals as in quantum mechanics or field theory.

One way to think about the mutual information is as a measure of dependency or correlation.  Clearly we have $I(x;y) = 0$ if the variables $x$ and $y$ are independent, so that $P(x,y) = P_X(x)P_Y(y)$, and Shannon's arguments tell us that $I(x;y)$ is the only measure of correlation that satisfies a set of sensible requirements.  If $x$ and $y$ are jointly Gaussian, then 
\begin{equation}
I(x;y) = -\frac{1}{2}\log_2(1-c^2),
\end{equation}
where $c$ is the usual linear correlation coefficient. But it is easy to imagine relations between $x$ and that are highly informative but not linear.  Even if the underlying variables are linearly correlated, the things we measure might be nonlinearly transformed version of these variables, and in this case computing the mutual information becomes an especially powerful way of analyzing the data; see Section \ref{parameters} below.

The more conventional way of thinking about the mutual information is in terms of a communication channel.  There is a input message $x$, and it is sent along channel that produces the output $y$.  The fact that $y$ is chosen from a probability distribution [$P(y|x) = P(x|y)P_Y(y)/P_X(x)$] means that the communication is noisy, and this must limit the amount of information that can be transmitted.  If we think of the mutual information as being a functional of the distributions $P_X(x)$ and $P(y|x)$, then the convexity of the logarithm means that there is a maximum as functional of the distribution of inputs; this maximum is called the channel capacity,\footnote{It is a remarkable fact that one can transmit information {\em without error} along a channel that has a non--zero noise level, provided that the rate of transmission does not exceed this capacity.  Strictly speaking ``capacity'' usually is used to describe this maximum rate of errorless transmission, rather than the maximum of the mutual information between input and output \cite{cover+thomas_91}.} and is the limit to information transmission set by noise in the channel.\footnote{In engineered systems, much emphasis is placed on devising coding schemes, i.e., algorithms that transform inputs $x$ into messages to be sent through the channel (and likewise recover $x$ from the channel output), so that information can be transmitted over noisy channels as close to capacity as possible using bounded processing resources. While interesting, coding questions have remained largely unexplored in biological systems.} 

We note that actually finding the capacity of a channel, given a detailed model of the input/output relation and noise as represented by $P(y|x)$, can be challenging.  Simple models, such as $y$ being a linearly filtered version of $x$ with added Gaussian noise, are tractable, and generalize in a straightforward fashion to cases where the signals are varying in time. For this reason, Gaussian channels have dominated the discussion of both biological and engineered systems, and the capacity-achieving strategies that depend on the system's input and output noise levels are well-understood \cite{bialek_12,cover+thomas_91,berger_71}.  Beyond the Gaussian channel, progress has been made using various approximative schemes, such as by assuming that noise is small compared to the dynamic range of the signal (the small-noise approximation), but there is certainly room for further theoretical developments using biophysically-informed noise models.

Examples of information transmission through a noisy channel are everywhere in biological systems.  Most obviously, our eyes and ears take inputs from the outside world and the neurons emerging from our sense organs generate streams of identical electrical pulses, termed action potentials or spikes.  Even single celled organisms have sensors,  as with the bacterial chemotaxis system, which responds to changing concentrations of molecules in the environment and generates at the output changes in the direction of swimming.  There are also purely internal examples, as when a cell synthesizes ``transcription factor'' proteins that bind to DNA and influence the rate at which encoded information is read out to make other proteins; here the transcription factor concentration is the input and the resulting protein concentration is the output.  In some of those cases variables are naturally continuous (sound pressure), in other cases they are naturally discrete (counting individual molecules), and in many cases it is useful to pass between descriptions that are either discrete or continuous (e.g, concentration instead of counting molecules); it is attractive that we can discuss information transmission in all these cases, in the same language and the same units.  

Information is a difference in entropies, as in Eq (\ref{IdiffS1}), and because information is mutual we can write this in two ways,
\begin{eqnarray}
I(x;y) &=& S[P_Y(y)] - \langle S[P(y|x)]\rangle_{P_X(x)} \label{info2} \\
&=& S[P_X(x)] - \langle S[P(x|y)]\rangle_{P_Y(y)}, \label{info3}
\end{eqnarray}
where $\langle\cdot\rangle_{p(x)}$ denotes an average over $p(x)$.  If either the input or the output is discrete, then both the entropy and the conditional entropy are positive semi--definite, and hence the entropy sets an upper bound on the information.  Further, if we can assign a cost to every output state $y$---e.g., because generating more action potentials costs more energy, then there is a maximum entropy consistent with a given average cost (for more on maximum entropy see Section \ref{maxent}).  Thus there is a rigorous path from limits on physical resources to limits on information transmission.
 
Another way of bounding information is through the data processing inequality.  If the output $y$ receives information about the input $x$ only through some intermediate $z$, we can write 
\begin{equation}
P(y|x) = \sum_z P_{\rm II} (y|z) P_{\rm I} (z|x) .
\end{equation}
Now there are two communication channels, first $x\rightarrow z$ (channel $\rm I$) and then $z\rightarrow y$ (channel $\rm II$).   Each channel is associated with a mutual information, $I_{\rm I}(x; z)$ and $I_{\rm II} (z; y)$, and each of these is an upper bound on the information transmission through the combined channel, $I(x; y)$.  Another way of saying this is that any processing  from $z$ to $y$ can only result in a  loss of information, so that $I(x;y) \leq I_{\rm I} (x;z)$.   In this sense, information can be destroyed, but not created.  This result is conceptually important, and we will see that it is of practical use in analyzing experiments (Section \ref{parameters}).  As for the mutual information, so too for the channel capacities:  to support mutual information $I(x;y)$, between $x$ and $y$, the capacities of both channels $\rm I$ and $\rm II$ must be larger than $I(x;y)$.

The universality of information-theoretic quantities is their strength, but it also leaves one with an uncomfortable feeling that there is more to the story of biological signal processing than bits. In particular, not all bits may be of equal value or utility to the organism.   Information processing usually has a goal beyond communication, and this seems especially true in biological systems.  Perhaps because Shannon called his work a mathematical theory of communication, it is not as widely appreciated that information theory has anything to say about such goal--defined tasks.  The key idea is rate--distortion theory, which again traces back to Shannon \cite{shannon_59,berger_71}.

As an example, consider a single celled organism in an environment where various nutrients are present at concentrations ${\mathbf c} = \{c_1,\, c_2,\, \cdots ,\, c_N\}$, and over time this environmental state varies.  To metabolize these molecules, the cell synthesizes $e_1$ molecules of enzyme $1$, $e_2$ copies of enzyme $2$, and so on up to enzyme $K$; to fully convert the inputs into useful molecules, there may need to be more enzymes than nutrients.  The growth rate of the cell depends on both the environment and the state of the cell, $\lambda ({\mathbf c},{\mathbf e})$.  Under any particular environmental condition $\mathbf c$, there is an optimal setting of the internal state $\mathbf e$ that maximizes the growth rate.  If the cell finds this optimum exactly, then by looking at $\mathbf e$ we would know exactly the environmental variable $\mathbf c$, and this exact knowledge of real variables corresponds to an infinite amount of information.  At the opposite extreme, the cell could choose the copy numbers of various enzymes at random, and hope that, averaged over the distribution of environments, the average growth rate is not so bad; in this case the internal state of the cell carries zero information about the environment.    But neither infinite nor zero information seems plausible; what happens in between?  Rate--distortion theory tells us that if the cell needs to achieve a given average growth rate, then there is a minimum required mutual information between the state of the environment and the internal state of the cell, $I ({\mathbf c},{\mathbf e}) = I_{\rm min} (\langle\lambda\rangle )$.  

More generally, if we have inputs $x$ and outputs $y$, then given some measure of performance ${\cal U}(x,y)$, we can plot the average performance vs. the mutual information $I(x;y)$.  The rate--distortion curve $I_{\rm min} (\langle {\cal U}\rangle )$ divides this plane into accessible and inaccessible regions.  Given a certain number of bits, there is a maximal achievable performance, and conversely achieving any given level of performance requires a minimum number of bits.  Thus, although collecting and transmitting bits cannot be the goal of an organism, any real goal requires a minimum number of bits.  If bits are cheap, then this is a true but not very useful observation.  If, on the other hand, the limited resources available to cells and organisms mean that more bits are hard to get, then selection for better performance will translate into selection for mechanisms that gather and process more bits with the same resources.  

Rate distortion is a framework that formalizes lossy data compression: signals $x$ can be sent through the channel to yield outputs $y$ which are then used to find the best achievable reconstruction $\hat{x}(y)$ of the input signal. The metric for the reconstruction quality is the distortion function, for example, the RMS error between the true signal $x$ and its reconstruction $\hat{x}$. In this sense, rate distortion tells us the minimum number of bits to recover the input signal to within a desired level of quality. When thinking about biological systems, it is often convenient to turn this reasoning around and explicitly construct various decoding or readout schemes that reconstruct the input signals from the activity of the biological network. Classic examples involve decoding of motion trajectories from the spiking activity of  neurons in the visual system \cite{bialek+al_91,marre+al_13}, or the decoding of hand trajectories  from the activity of neurons in motor cortex during drawing \cite{schwartz_94}; for more on decoding of neural activity see Ref \cite{spikes}.  Decoding schemes can be used to put a lower bound on information transmission (Section~\ref{sec3}) and might further suggest how much of the total information is encoded explicitly, in a way that can be extracted using biologically plausible mechanisms. Indeed, the relationship between the total information and the component which is accessible in simple form is an issue in neural coding that needs to be explored more fully.

\section{Information and data analysis}
\label{sec3}

In complex systems, we may be unsure which features of the system dynamics are the most relevant.   Even if we can identify the correct features, we might not know the biologically meaningful metric that measures the similarity or difference along these relevant dimensions.  Information theory gives us a way out of these difficulties, because it singles out measures that are completely general:  while some particular correlation function of $X$ and $Y$ might vanish, if these variables are related in any way then there must be a nonzero mutual information $I(X;Y)$; further, there is an obvious sense in which the stronger the relationship, the greater the mutual information.  Against this generality is the difficulty of actually estimating the entropy of a probability distribution from a limited set of samples.

\subsection{Quantifying real information flows}
\label{measure_info}

In thermodynamics, entropy changes are related to heat flows, and this can be used to measure the entropy in equilibrium systems.  Away from equilibrium statistical mechanics, we have no such general method for measuring entropy.  Indeed, the entropy depends on the entire probability distribution, and so ``measuring'' the entropy is quite unlike measuring the usual observables.  

We can estimate the entropy of a probability distribution by counting how often each state occurs, using this frequency as an estimate of the probability, and plugging into the definition $S = -\sum_{\rm n}p_{\rm n} \log_2 p_{\rm n}$.  While the errors in estimating the individual probabilities $p_{\rm n}$  are random, the convexity of the logarithm means  that these random errors contribute to a systematic error than declines as $1/\sqrt{N_s}$, where $N_s$ is the number of samples.  The difficulty is that the coefficient of this systematic error is proportional to the number of possible states of the system, and hence accurate estimates require many more samples than states.  The problem becomes worse when we try to estimate the mutual information between two variables, since the number of possible states is the product of the number of states for each variable taken separately.  

The problem of systematic errors in estimating entropy and information was appreciated very early, in the 1950s, as information theoretic ideas began to be applied to biological problems \cite{miller_55}. Motivated in large part by efforts to measure the information that neural responses carry about sensory inputs, there has been a substantial effort to calculate and correct for these systematic errors \cite{panzeri+treves_95,strong+al_98,paninski_03}.   A different direction is to note that the single number, entropy, might be well determined even if the full distribution is not.  This idea can be made concrete in a Bayesian approach, placing a prior on the space of probability distributions; priors that are uniform in entropy, as opposed to uniform in probability, seem especially effective \cite{nemenman+al_02,nemenman+al_04,archer+pillow_12,archer+pillow_13}.  In many cases, the variables of interest are not naturally discrete, and so the problem of entropy estimation is tangled with that of density estimation \cite{victor_02}.  Although much progress has been made, the difficulty of estimating entropy and information from finite samples leads to periodic inventions of other measures that might serve similar purposes (e.g., quantifying dependency beyond linear correlation); for a recent reminder of why mutual information is not just one measure among many, see Ref \cite{kinney+atwal_14}.

An important ingredient in estimating entropy and information from real data has been the use of bounds.  It may be difficult to estimate the entire probability distribution, but it often is feasible to measure, with small error bars, several average quantities (e.g., moments), and construct the probability distribution that matches these averages exactly but has the maximum possible entropy (Section \ref{maxent}).  The data processing inequality means, for example, that if we want to measure the mutual information between $x$ and $y$, and $y$ is of high dimensionality, any projection into a lower dimensional space, $y\rightarrow z$, can only result in a loss of information, $I(z; x) \leq I(y;x)$.  Reducing dimensionality makes the problem of estimating distributions much easier, and thereby eases the sampling problems of entropy estimation; since the lower dimensional variable provides a lower bound to the true information, it makes sense to search for optimal projections (Section \ref{parameters}).  Finally, as emphasized long ago by Ma \cite{ma_81}, the probability that two randomly chosen states are the same, $P_c = \sum_{\rm n} p_{\rm n}^2$, provides a lower bound to the entropy, $S \geq -\log_2 P_c$, that can be estimated reliably from relatively few samples.  For more details, see Appendix A.8 of Ref \cite{bialek_12}.

 A modest though widespread use of information theory in the analysis of biological data concerns the description of regulatory sites along DNA.  The readout of information encoded in the genome---the synthesis of proteins---is regulated in part by the binding of ``transcription factor'' proteins (TFs) to these sites, and the strength with which one TF binds depends on the DNA sequence at the binding site. In this sense, the local sequences can be thought of as ``addresses'' that distinguish certain genes from the rest, but it is known that transcription factors do not do exact string matching; instead, they bind to an ensemble of sites that permit some variation.  Early work used simple arguments to compute how much ``information'' is contained in these binding site ensembles, i.e., as the difference between the entropy of random sequences and the difference of the ensemble that the TF will bind \cite{berg+hippel}, and this has given rise to a popular graphical way of depicting TF binding site specificity in the form of ``sequence logos'' \cite{seqlogo}. Subsequent work has refined these arguments and asked whether the information in the binding sites is sufficient to address specific genes in genomes of pro- and eukaryotes \cite{wunderlich+mirny}. 

Meaningful quantitative statements about information flow obviously depend on having high quality data.  In the nervous system, the fact that the relevant output signals are electrical means that quantitative measurements have been the norm since the invention of the vacuum tube \cite{adrian_28}.  Steady improvements in technology now make it possible to collect sufficient data for information theoretic analyses in almost any neural system, even in freely behaving animals; in practice, much of the motivation for these analyses has come from ideas about the optimization of the neural code, and so these measurements are reviewed below (Section IV.B).  In biochemical and genetic networks, the fact that outputs are the concentrations of molecules means that there is no universal measurement tool.  

Classical methods for measuring the concentrations of signaling and control molecules inside cells include raising antibodies against particular molecules of interest, and then tagging antibodies against these antibodies with fluorescent labels.  Recent work in the fly embryo shows that, with care, this classical technique can pushed to give measurements of concentration that are accurate to $\sim 3\%$ of the relevant dynamic range \cite{dubuis+al_13a}.  More modern methods involve genetically engineering the organism to produce  fusion of the protein of interest with a fluorescent protein \cite{tsien_98}, and here too there has been progress in demonstrating that the fusion proteins can replace the function of the native protein, quantitatively \cite{gregor+al_07a,morrison+al_12}.  Finally, rather than monitoring protein concentrations, one can count the messenger RNA molecules (mRNA) that are an intermediate step in protein synthesis; recent work has pushed to the point of single molecule sensitivity, so that counting is literal \cite{raj+al_08,little+al_13}. Similarly, progress has been made in quantifying the activity of intracellular signaling networks, and subsequently using information theoretic approaches to compare how signals propagate in different network architectures \cite{cheong+al_11}, or how they are encoded in the dynamical properties of the response \cite{selimkhanov+al_14}.

\subsection{Quantifying correlations}

In the last decade it has become possible to perform simultaneous recordings from tens or hundreds of nodes in interacting biological interacting networks. Gene expression arrays and sequencing technology have enabled such readouts for gene expression patterns; cheap sequencing can map the distributions of gene variants in populations; multi-electrode arrays and lately optical imaging have pushed the boundaries in neuroscience to the point where full temporal activity traces for every neuron in small brains---a zebrafish larva, the worm {\em C elegans}, or the fruit fly---don't seem out of reach; and in the field of collective behavior, simultaneous tracing of positions of hundreds or thousands of individuals can  be performed.  Because the barriers between subfields of biology are large, each of these experimental developments has led to an independent discovery of the need for new and more powerful analysis tools.  Even the simple question of when the activity at two nodes in the network is correlated is complicated, because it is surely not true that linear correlations capture the full relationships among pairs of variables in these complex systems.

As microarray methods provided the first global pictures of gene expression in cells, several groups tried to analyze the covariation in expression levels simple by computing correlations.  With enough data, it became possible to estimate the mutual information between expression levels \cite{slonim+al_05a}.  These initial analyses revealed that, as expected, linear correlations did not exhaust the relationships between genes; while the most correlated pairs had the highest mutual information, at intermediate or even minimal levels of correlation one could find pairs of genes whose expression levels shared nearly half the mutual information of the most correlated pairs.  In addition, it became clear that pairs of expression levels could share more than one bit of mutual information, which contributed to the realization that the control of gene expression involved more than just on/off switches.    

Measurements of mutual information also provide a new basis for clustering, in which we try to maximize the average mutual information between the expression levels of pairs of genes in the same cluster \cite{slonim_05}.  This approach is interesting mathematically because it is invariant to any invertible transformations of the expression levels (and hence to systematic distortions in the measurement), and because the clusters have no ``centers'' or prototypical members.  These information theoretic methods were able to reveal biologically significant structures even inside tightly correlated ``modules'' where other methods failed.  Further developments place the estimates of mutual information among many pairs of gene expression levels at the core of algorithms for reconstructing the underlying network of interactions, where the data processing inequality provides a guide to disentangling direct interactions from more widespread correlations \cite{aracne}.

One can look at interactions between the genes in a very different way, analyzing joint presence or absence of particular genes in the sequences of many organisms. Here, each ``sample,'' i.e., one organism's genome, is represented as a long binary vector, where 1 represents a presence of a particular gene  and 0 its absence.\footnote{Some care is needed here.  Two different organisms almost never have genes with exactly the same sequences, but if we are comparing organisms that are not too far separated in evolutionary distance, then finding the corresponding or ``homologous'' genes is relatively straightforward.  It is more difficult to be certain of the historical path that generated these homologs:  descent from a common ancestor (orthologs'), convergent evolution, gene duplication (paralogs), ... .  For more about sequence similarity, see Section \ref{maxent}.} Estimating the mutual information between the presence/absence of pairs of genes is technically straightforward, and reveals the organization of genes into clusters \cite{bowers};  this approach can be extended beyond pairs to probe higher--order interactions among  genes. Information theoretic measures can also be used to quantify the correlations between the presence/absence of particular genes and various phenotypes, and these analyses point toward  functional modularity in the organization of genomes \cite{slonim+elemento}.

\subsection{Systems with many degrees of freedom}
\label{maxent}

One of the great triumphs of the twentieth century is the identification and characterization of the molecular building blocks of life.  But life is more than the sum of its parts. From the spectacular aerial displays by flocks of birds down to the coordinated movements of cells in a gastrulating embryo,  many of the most striking phenomena in the living world are the result of interactions among hundreds, thousands, or even millions of individual elements.  The enormous success of statistical physics in describing emergent phenomena in equilibrium systems has led many people to hope that it could provide a useful language for describing emergence in biological systems as well.    Perhaps the best developed set of ideas in this direction concerns the dynamics of neural networks \cite{hopfield_82,amit_89,hertz_91}, but there have also been important developments in the description of collective animal behavior \cite{toner+tu_95,toner+tu_98,ramaswamy_10} and in models for genetic networks.  While each case has its own special features, it seems fair to say that, in all these cases, the search for simple, analyzable models often takes us away from contact with the details of measurements on the real biological systems.  In extreme cases, it is not clear whether these models are just metaphors (the alignment of birds in a flock is like the alignment of spins in a magnet), or if they are to be taken seriously as theories that make quantitative predictions.  Over the past decade, progress in experimental techniques, as noted in the previous section, have held out the promise of much more detailed confrontations between theory and experiment.  

Although there are many interesting questions about emergent behavior in biological networks, we focus here on one question where information theoretic ideas have been crucial:  if there a variable $y_{\rm i}$ that lives at each node of the network, can we make a model for the joint distribution of these variables, $P_y({\mathbf y}) \equiv P_y(y_1 ,\, y_2, \, \cdots ,\, y_N)$?  Progress on this problem has been made using the idea of maximum entropy \cite{jaynes_57}.

Let us assume that experiments provide $T$ observations on the state of the system, $\mathbf{y}$. Often, the dimension of the system is so large that $T$ samples are completely insufficient to empirically sample $P_y(\mathbf{y})$. However, we might be able to estimate certain statistics on the data, which we denote by $\langle \hat{O}_\mu\rangle_{\rm expt}$, where $\mu=1,\dots,M$. These could be, for example, expectation values of individual network nodes, $\langle y_i\rangle$, the covariances, $\langle y_iy_j\rangle$, or any other function for which the samples provide a reliable estimate. We look for a model distribution $\hat{P}_y(\mathbf{y})$ that will exactly reproduce the measured expectation values, but will otherwise be as unstructured, or random, as possible---hence maximum entropy. This amounts to solving a variational problem for $\hat{P}$, for which the solution is:
\begin{equation}
\hat{P}^{(\hat{O}_1,\dots,\hat{O}_M)}(\mathbf{y})=\frac{1}{Z(\{g_\mu\})}\exp\left[\sum_{\mu=1}^M g_\mu \hat{O}_\mu(\mathbf{y})\right]. \label{maxent1}
\end{equation}
Thus, the maximum entropy distribution has an exponential form, where the coupling constants $g_\mu$ have to be chosen so that the constraints $\langle \hat{O}_\mu\rangle_{\rm expt}=\langle \hat{O}_\mu\rangle_{\rm \hat{P}}$ are exactly satisfied.   Equation (\ref{maxent1}) has the form of a Boltzman distribution in which every state of the system is assigned an ``energy'' $E({\mathbf y}) = \sum_{\mu=1}^M g_\mu \hat{O}_\mu(\mathbf{y})$, and this equivalence to equilibrium statistical mechanics is an important source of intuition and calculational tools.  But Eq (\ref{maxent1}) is {\em not} a description of a system in thermal equilibrium.  If the only thing we measure about a system is its energy, then asking for the maximum entropy distribution consistent with the average energy does indeed correspond to a description of thermal equilibrium, but in most biological systems the quantities that we measure have no relation to the energy, and we are measuring many of them.   

The maximum entropy method does not correspond to a single model for the system, but rather provides a systematic strategy for building a hierarchy of models that provide increasingly good approximations to the true distribution.  At each step in the hierarchy, the models are parsimonious, having the minimal structure required to reproduce the expectation values that we are trying to match.  For every new expectation value that we can measure, the new maximum entropy model provides a tighter upper bound on the entropy of the true distribution, and can uniquely decompose the total amount of correlation among the $N$ nodes in the network into contributions from pairwise, triplet-, $\dots$, $K$-fold interactions \cite{schneidman+al_03}; the hope is that we arrive at a good approximation while $K\ll N$. 

Equation (\ref{maxent1}) tells us the form of the maximum entropy distribution, but to complete the construction of the model we need to find the coupling constants $\{g_\mu\}$.  Usually in statistical mechanics we are given the coupling and try to compute the expectation values or correlation functions ($\{g_\mu\} \rightarrow \{\langle{\hat O}_\mu ({\mathbf y})\rangle\}$).  Here we need to do inverse statistical mechanics, mapping the experimentally measured expectation values back to the coupling constants ($\{\langle{\hat O}_\mu ({\mathbf y})\rangle\} \rightarrow  \{g_\mu\}$).  Such inverse problems have a history dating back (at least) to the demonstration by Keller and Zumino \cite{keller+zumino_59} that  the temperature dependence of the second virial coefficient in a classical gas determines the interaction potential between molecules uniquely, provided that this potential is monotonic; the search for  rigorous statements about inverse statistical mechanics has  attracted the attention of mathematical physicists \cite{chayes+al_84,caglioti+al_06}, although many questions are open.  At a practical level, estimating the coupling constants from measured expectation values can be seen as part of the  broader problem of learning probabilistic models from data, and hence part of machine learning in computer science.  For a small sampling of the interplay between ideas in statistical physics and machine learning, see Refs \cite{mezard+montanari_09,tishby+al_90,yedidia+al_03,mezard_03,mehta+schwab_14}.

Interest in maximum entropy approaches to biological networks was stimulated by results from an analysis of activity patterns in small groups of neurons from the vertebrate retina \cite{schneidman+al_06}.  As the retina responds to natural movies, pairwise correlations are weak, of mixed signs, and widespread.  Models that ignore these weak correlations make predictions about the frequency of patterns in groups of ten neurons that can be wrong by many orders of magnitude, but these errors are largely cured in maximum entropy models that match the pairwise correlations.   Even in these small networks, there are hints that what one sees in the network are genuinely collective states.  By now, these ideas have been extended and applied to several different neural systems \cite{tkacik+al_06,shlens+al_06,tang+al_08,tkacik+al_09,shlens+al_09,ohiorhenuan+al_10,ganmor+al_11,simplest,sdme,tkacik+al_14a,mora+al_14}.  Most recently it has been possible to build models for the joint of activity of more than 100 neurons in the retina, and these models are so accurate that they can predict distribution of the effective energy deep into the tail of patterns that occur only once in a two hour long experiment \cite{tkacik+al_14a}.  The states that dominate the distribution are indeed collective, so one can reconstruct the time dependent response of single cells from the behavior of the population, without reference to the visual stimulus.  Most provocatively, that distribution that we reconstruct from the measured expectation values exhibits many of the signature of an equilibrium statistical mechanics problem near its critical point \cite{tkacik+al_14b}.

An independent stream of work has used maximum entropy ideas to think about the ensemble of amino acid sequences that form a family of proteins, folding into essentially the same three--dimensional structures.  This work was inspired by experiments that actively explored sequence space, trying to respect pairwise correlations among amino acids at different sites \cite{socolich+al_05,russ+al_05}, using an algorithm that is equivalent in some limits to maximum entropy \cite{bialek+ranganathan_07}.  Although these correlations  extend over long distances, the effective interactions among amino acids in these models were found to be spatially local;  this raises the possibility that we can infer the spatial neighbor relations among amino acids from data on sequence variation, in effect folding the protein via sequence comparisons  \cite{weigt_al_09,marks+al_11,hopf+al_12,sulkowska+al_12}.    This new approach to the protein folding problem has generated considerable excitement, but there are many open questions about the sufficiency of the available data and the approximations that have been used to solve the inverse problem.  Note that this effort to describe the distribution of sequences consistent with a given protein structure  is the inverse of the usual protein folding problem, where we are given the sequence and asked to predict the structure, and there is a big conceptual question about the relation between these two problems.

In the immune system we can look at special cases of sequences ensembles, the distribution of antibody molecules that organisms use to combat infection and the distribution of viral sequences in patients that have long term infections such at HIV.  In the problem of antibody diversity, maximum entropy models built from pairwise correlations provide a surprisingly accurate description of the distribution as a whole \cite{mora+al_10}, while for HIV the effective energy in maximum entropy models  predicts viral fitness and vulnerability to immune attack \cite{ferguson+al_13}.   This is a small window into a much larger literature on statistical physics approaches to the immune system.\footnote{See, for example,  the program at the Kavli Institute for Theoretical Physics, http://immuno-m12.wikispaces.com.}

Birds in a flock are thought to follow only  their near neighbors, yet the entire flock can decide to move in a single direction and at nearly uniform speed.  Order spreads through the flock much as through spins in a magnet, and   this is more than a metaphor:  maximum entropy models consistent with measured local correlations in a flock provide successful quantitative, parameter--free predictions of correlations in flight direction and speed throughout the flock \cite{bialek+al_12,bialek+al_14}.  It is also possible to do this analysis in more detail, building maximum entropy models that match the correlations between a bird and its nearest neighbor, its second neighbor, and so on; one finds that the resulting couplings which link each bird to its $k^{\rm th}$ neighbor decline exponentially with $k$, providing some of the most direct support yet for the hypothesis that interactions are local \cite{cavagna+al_14a}.  Real starling flocks exhibit long--ranged correlations of the fluctuations in flight direction and speed \cite{cavagna+al_10}; analysis of the maximum entropy models shows that the directional correlations arise from Goldstone modes associated with the spontaneous polarization of the flock \cite{bialek+al_12}, while the speed correlations are long--ranged only because the parameters of the model are tuned close to a critical point \cite{bialek+al_14}; signatures of criticality have also been found in the behavior of insect swarms \cite{attanasi+al_14}.

\subsection{Information-theoretic parameter inference}
\label{parameters}

A standard tool for fitting parametric models to data is maximum likelihood inference. This procedure, however, depends on our ability to write down the likelihood function: the probability of making specific experimental observations given the parameters of the model. In studying biological systems we often are  faced with situations where this is impossible, for two very distinct reasons. First, while we might have a good idea for how a part of the system works (e.g., that transcription factors recognize and bind DNA sequences in a way that is governed by equilibrium statistical mechanics), what we can measure is far ``downstream'' of these events  (the activation level of a gene), and we may not have a quantitative model for the intervening steps.    Second, and less fundamentally, many experiments still are not as calibrated as one might hope, so we don't have a model for the noise in the measurement itself.    

Without the ability to write down the likelihood function, is there any way to do unbiased parametric inference of the process of interest?  There is isn't much we can do about the general version of this problem, but in many biological systems one crucial step in the process that we are trying to describe involves a substantial reduction of dimensionality.  Thus, in the visual system, we expect that neurons respond not to every possible detail of the image or movie falling onto the retina, but only to some limited set of features; in the simplest case these features might be described simply as linear projections onto a template or ``receptive field.''  If this picture is right, then the transformation from input movie $x$ to output spikes $z$ can be written as $x\rightarrow  y \rightarrow z$, where $x$ is a vector in a space of very large dimensionality (the space of movies), while both $y$ and $z$ are low dimensional, in the simplest case just scalars. The mapping $y\rightarrow z$ then is easy to characterize; the problem is how to choose the projection that describes $x\rightarrow y$.  The data processing inequality tells us that the best possible predictions of $z$ will be generated by a model in which the mapping $x\rightarrow y$ captures as much information as possible about $z$.  Importantly, the difficulty of searching for such ``maximally informative projections'' does not vary much with the structure of the distribution of the inputs $x$, which means that we can use this approach to characterize the encoding of complex, naturalistic input signals \cite{sharpee+al_04}.   Applications and extensions of this idea have been explored in several different systems \cite{sharpee+al_06,fitzgerald+al_11,rajan+bialek_12,eickenberg+al_12,rajan+al_13}.

The idea of searching for maximally informative projections has also been used to analyze experiments on the interactions of transcription factors with their target sites along DNA; here the projection is from the local DNA sequence to the binding energy of the TF \cite{kinney+al_07}.  An important success of this initial effort was to show that two independent experiments which characterized the same transcription factor in fact led to consistent models of TF--DNA binding, despite earlier claims to the contrary.   This work also demonstrated the formal equivalence between information-theoretic parameter inference and ``error-model-averaged'' maximum likelihood inference.   Subsequent experiments exploited these idea to give a much more detailed view of how a transcription factor and the RNA polymerase bind and interact at the lac operon \cite{kinney+al_10}, and related methods were developed to allow analysis of a wider range of input data \cite{elemento+al_07}.

Because the issues of model inference for processes embedded into unknown larger systems or subject to unknown noise processes are prevalent in biology, we expect to see further applications for information theoretic inference in biology. We also note that these applications can be seen as specific examples of a more general framework that relates learning (including parameter inference) to information flow between the data and its compressed representations. There are many interesting links between the learning problem, statistical mechanics, and field theory, which are beyond the scope of this review.

\section{Optimizing information flow}
\label{sec5}

Information is essential for life, but bits are not free.  Is it then possible that organisms are driven to extract and represent the maximum possible information given the physical constraints?  This is an idea that arose almost immediately after Shannon's original work, and has been a productive source for thinking about many specific biological systems, across many scales.

\subsection{The genetic code}

Five years after Shannon's papers, Watson and Crick proposed the double helical structure for DNA, and after another five years Crick could articulate ``the sequence hypothesis'': the sequence of amino acids in proteins is determined by the sequence of bases along the DNA \cite{crick_58}. Once one has this idea, the mapping between two major classes of biopolymers becomes a problem of coding, and it seemed natural to bring the new ideas of information theory to bear on this central  problem of life \cite{crick_63}.  In particular, it was hoped that the code might be efficient in the sense defined by Shannon, using as few bases as possible to encode the amino acid sequence of a protein.  As we now know,  the genetic code maps triplets of bases (codons) into one of twenty amino acids (plus ``stop''), so that there is substantial redundancy.   

The failure of the real genetic code to instantiate any of the early ideas about efficiency or optimality must have been a disappointment, and it is hard to find much serious interaction between information theory and molecular biology for some time after the code was worked out.   This split may have been premature.  In a triplet code, there are three possible ``reading frames,'' and the first organism to have its entire genome sequenced---the bacteriophage $\Phi$X174---makes use of the same DNA in multiple reading frames \cite{sanger+al_77}, an unexpected multiplexing that enhances efficiency in the original Shannon sense.  Another observation is that the redundancy of the code is structured, so that random changes of a single base along the DNA are likely to map one amino acid either into itself (`silent' mutations) or into amino acids with very similar chemical properties.  This would not be true in a random assignment of codons to amino acids, and it has been suggested that the real code may even be nearly optimal in this respect \cite{freeland+hurst_98,freeland+al_00}; if correct this means that the genetic code is not an optimal noiseless compression, as in the earliest (incorrect) proposals,  but may be near the bounds set by rate--distortion theory \cite{tlusty_08}.  A full development of this idea will require understanding the cost of amino acid substitutions, which relates back to the analysis of sequence ensembles in Section \ref{maxent}.  

The genetic code really is  a strategy for communicating from DNA bases to amino acids, and here too there are errors.  Although one often pictures the specificity of life's mechanisms as deriving from the specific binding between complementary molecular components (as in the pairing of A with T and C with G in DNA), a beautiful chapter in the physics of life is the appreciation that these equilibrium interactions are not sufficiently selective, and hence that cells must build Maxwell demons to sort molecules with higher precision; the broad class of mechanisms that implement these demons is called kinetic proofreading \cite{hopfield_74,ninio_75} [see also Section 4.5 of Ref \cite{bialek_12}].  This creates an interesting tradeoff for the cell, which can operate at high fidelity but high energetic cost (paying for the demon), or trade the cost of errors against the reduced energetic cost of correcting them \cite{ehrenberg+kurland_84}.    Although the costs and benefits of information flow in reading the genetic code have been much discussed, this still has not been given a fully information theoretic formulation.

\subsection{Efficient representations in sensory processing}

One of the earliest uses of information theory to think about biological systems was an attempt to estimate the amount of information carried by neuronal action potentials \cite{mackay+mcculloch_52,spikes}.  The answer depends crucially on the time resolution with which such spikes can be generated and detected by neurons.  If the average rate of action potentials is $\bar r$, and the system can work with a time resolution $\Delta\tau \ll 1/\bar r$, then the stream of spikes becomes a binary sequence of $1/\Delta\tau$ symbols per second, and the probability of a 1 (spike) is ${\bar r}\Delta\tau$.  The maximum entropy of this sequence, which sets the capacity for the neuron to carry information, is then $S\sim \log_2(e/\bar r \Delta\tau)$ bits/spike.  In contrast, if the system is limited to counting spikes in windows much larger than the typical interspike interval, then the entropy becomes $S\sim \log_2 (e\bar r\Delta\tau)/(\bar r\Delta\tau)$ bits/spike.  Evidently higher time resolution allows for higher information rates, and dramatically so:  if spike rates are a few per second (as often true, on average), then time resolutions of a few milliseconds could allow for up to ten bits per spike.  In 1952, when MacKay and McCulloch first made such estimates, locating real neural codes on the continuum from timing to counting was a matter of speculation.  But this work established an agenda:  could we measure the information that spike trains carry, for example about sensory inputs, and does this information approach the limits set by the entropy of the spike sequences themselves?

MacKay and McCulloch were thinking about a single neuron.  But signals often are conveyed by larger populations of neurons, for example in the retina where many neurons respond to overlapping regions of the visual world.  As early as 1959, Barlow suggested that processing in the retina would have the function of removing redundancy among the outputs of neighboring ganglion cells, and argued that several major features of these cells' responses could be understood in this way: a reduction in the response to sustained stimuli (termed adaptation) serves to reduce redundancy in time, while the tendency of cells to respond to differences between light intensity in a small patch and the surrounding region (``center--surround'' organization) serves to reduce redundancy in space \cite{barlow_59,barlow_61}.

It would take many years until these ideas about efficient coding in sensory processing would be made quantitative.  An important qualitative idea, however, is that the efficiency of a code depends on the distribution of inputs.  Thus, we might expect that real organisms use codes that are matched to the distribution of signals that they encounter in their natural environment; put simply, the function of the code would make sense only in context, and the relevant context is quite complex.  This realization stands in contrast to the long tradition of using highly simplified stimuli in laboratory studies of sensory function, and has led to the emergence of a subfield of neuroscience concerned specifically with the structure of natural sensory signals and the neural response to these signals \cite{simoncelli+olshausen_01,geisler_08}.

The first stages of retinal coding occur with neurons that produce analog voltages rather than discrete spikes.  Laughlin focused on cells in the first stage of processing after the photoreceptors in the fly retina, and asked if the nonlinear input/output relation that transforms light intensity into voltage is matched to the environmental distribution of light intensities \cite{laughlin_81}.  If the voltage noise is small and constant throughout the dynamic range, then optimal information transmission implies that a (normalized) input/output relation should be equal to the cumulative probability distribution of the light intensity.  By sampling natural scenes with a custom-built photodetector to predict the optimal input/output relation and comparing it with the measured one, Laughlin found a remarkable agreement with the theory,  especially considering that there are no free parameters. Although there are obvious open questions, this was a really beautiful result that inspired the community to take these ideas more seriously.

For spiking neurons, measurements of information transmission in response to complex, dynamic stimuli began to appear in the 1990s.  Initial measurements were based on decoding the spike trains to recover estimates of the input sensory signals \cite{bialek+al_91,rieke+al_93,borst+theunissen_99}, and then analysis tools for more direct estimates were developed \cite{strong+al_98,ruyter+al_97}.  In the primary sensory neurons of the frog inner ear, the mechanical sensors of the cockroach, ganglion cells in the vertebrate retina \cite{koch+al_06} and subsequent layers of visual processing in mammals \cite{reinagel+reid_00,liu+al_01}, and motion--sensitive neurons deeper in fly visual system, as well as other systems, it was found that spike trains could transmit more than one bit per spike, and typically reached 30--50\% of the limit set by the spike train entropy, even with time resolutions in the millisecond range. In accord with the  idea of matching to natural signals, information rates and coding efficiencies are larger with distributions of stimuli that capture features of the natural ensemble \cite{rieke+al_95}, and in some cases truly natural stimuli reveal codes that maintain their efficiency down to sub--millisecond resolution \cite{nemenman+al_08}.\footnote{These results depend on getting control over the problems of estimating entropy and information from limited samples, as discussed in Section \ref{measure_info}.} Despite much prejudice to the contrary, there are  demonstrations of efficient coding, with time resolutions of a few milliseconds, by single neurons in the cerebral cortex of our primate cousins \cite{buracas+al_98,kara+al_00}, suggesting strongly that these principles extend to brains much like our own.

If neurons implement coding strategies that are matched to the distribution of sensory inputs, what happens when this distribution changes?  When we walk from a dark room out into the bright sunlight, the mean light intensity changes, and it takes a moment for our visual system to adapt to this new level.  This adaptation is visible in the response of individual neurons in the retina, and this is one of the classical observations about neural coding, having roots in Adrian's first experiments in the 1920s \cite{adrian_28}.  But the efficient coding hypothesis predicts that the code should adapt not just to the mean, but to the whole distribution.  In particular, natural images have an intermittent structure, so that the variance in light intensity varies significantly from point to point in a scene, and hence the variance seen by neurons responding to a small patch of the world will vary in time, so adaptation to the variance would enhance efficiency \cite{ruderman+bialek_94}, and similar arguments could be made about other sensory modalities.   Early experiments in the vertebrate retina demonstrated this adaptation to variance explicitly \cite{smirnakis+al_97}.  Subsequent experiments have shown that retinal adaptation is an even richer phenomenon, so that properly chosen distributions of inputs can drastically change even very basic features of the code exhibited by single neurons, and this is part of a growing realization that the retina is capable of much more complex computations than originally believed \cite{gollisch+meister_10}.  

Adaptation to the variance  of input signals has now been seen in many systems \cite{wark+al_07}:   motion--sensitive neurons of the fly visual system \cite{brenner+al_00a,fairhall+al_01}, auditory neurons at several stages of processing in mammals and birds \cite{kvale+schreiner_04,dean+al_05,nagel+doupe_06,wen+al_09,dahmen+al_10,rabinowitz+al_11}, whisker motion sensing neurons in rodents \cite{maravall+al_07}, and in the farthest reaches of primate visual cortex, where cells responsive to the shapes of objects adapt to the variance along different parameters describing these shapes \cite{debaene+al_07}.  In many of these systems, changing the variance of the input distribution simply produces a rescaling of the input/output relation, as first observed in the fly motion--sensitive cells \cite{brenner+al_00a}.  If the typical signals are large compared to internal noise levels but not so large as to saturate the system, then the dynamic range of the inputs themselves provides the only scale in the problem, and hence optimization must lead to the observed scaling behavior;  in the fly it was even possible to check that the scaling factor chosen by the adaptation mechanism served to optimize the transmitted information \cite{brenner+al_00a}.  When the input distribution shifts suddenly, the input/output relation adjusts with remarkable speed, perhaps so rapidly that there is no sharp distinction between adaptation and the nonlinear transient response.  The time scales involved are essentially the shortest ones allowed by the need to gather statistics and verify that the distribution has, in fact, changed \cite{fairhall+al_01,wark+al_09}.

Barlow's original idea about redundancy reduction in the retina was implemented mathematically by Atick and Redlich \cite{atick+redlich_90}, who approximated retinal ganglion cells as analog devices that respond to a linearly filtered version of their visual inputs.  They searched for optimal filters, and showed that in low background lights ganglion cells should integrate over small regions of space and time, while at high light levels the response should be differentiating; this corresponds to a crossover between noise reduction and redundancy reduction being the dominant factors in enhancing information transmission, and agrees with the data.  

If we imagine that neurons filter their inputs to remove redundancy, then the outputs should become close to uncorrelated or ``white.''  But many natural input signals have power spectra in time that are approximately $1/\omega$, and whitening this  spectrum requires a transfer function between input and output $\sim \sqrt \omega$, which is not so easy to realize.  Nonetheless, the linear transfer function from the photoreceptor cells to the second order neurons studied by Laughlin has precisely this form \cite{laughlin+ruyter_96}.  The transfer of the signal from one cell to the next is mediated, here as elsewhere in the brain, by the neurotransmitter molecules that are packaged into discrete vesicles.  Careful measurements of the signal and noise characteristics in both cells leads to very high estimates of the information capacity of this connection (synapse), more than one thousand bits per second \cite{ruyter+laughlin_96}, and this is close to the physical limit set by the entropy of counting the vesicles \cite{spikes,bialek_12}.

In the actual retina, of course, not all cells are identical, and is interesting to ask if we can predict the distribution over cell types from optimization principles.   There is an old discussion about the choice of three color sensitivities for the photoreceptors in the bee's eye, arguing that the particular combination that is found in a whole family of related species serves to provide maximal information about the identity of the flowers from which bees collect nectar \cite{chittka+menzel_92}.  In the primate retina, more rigorous arguments have been used to predict the distribution of the three cone types, which is both highly non--uniform and varies with position on the retina \cite{garrigan+al_10}.  Retinal processing involves a plethora of cell types, but all visual systems divide signals into `on' and `off' pathways, which respond with more action potentials when the light level rises above or below its local average value; Ratliff et al \cite{ratliff+al_10} have argued that the preponderance of off cells is matched to an asymmetry in the distribution of light intensities found in natural scenes, and that other differences between on an off pathways also serve to optimize information transmission.  

If we focus on a single class of retinal ganglion cells, we find that they form a lattice across the surface of the retina [see, for example, Ref \cite{shlens+al_06}].  The geometrical properties of this lattice, specifically, the size over which each cell integrates the signal relative to the lattice spacing, can be predicted from information optimization \cite{borghuis+al_08}. These lattices are, however, not perfectly regular due to discreteness and imperfections in the spatial arrangement of the photoreceptors, and in principle these irregularities should lead to a loss of information.  Remarkably, these variations turn to be correlated with the local variations in the connectivity from the photoreceptor array to the ganglion cell layer in such a way as to almost perfectly compensate for the information loss \cite{liu+al_09}.

Attempts to analyze the optimization problem for spiking neurons were quite scattered until the work of Smith and Lewicki on the coding of complex sounds \cite{smith+lewicki_05,smith+lewicki_06}.  Rather than trying to build models of the process by which continuous inputs are converted into spikes, and then optimizing the parameters of these models, they chose the spike times themselves as the variables describing the encoding process  and optimized the code assuming that the goal was to reconstruct the inputs using a simple linear algorithm, following \cite{bialek+al_91}.  The result of the calculation is then a set of spike trains for a population of neurons responding to acoustic inputs, and these spike trains could be analyzed in the same way that experimentalists routinely analyze data from real neurons.  The results were impressive:  optimal spike encoders for speech--like sounds, but not for other sound ensembles, almost perfectly match the characteristics of real auditory neurons.  Further, the rate--distortion functions for these neural codes reach or even surpass the performance of the best artificial systems.  

The problem of optimal coding has also been studied from a theoretical angle. In a population of rate neurons that suffer from negligible intrinsic noise, the optimal code decorrelates the inputs: this should happen not only at pairwise order (by whitening), as suggested by early works of Barlow, van Hateren and others \cite{barlow_59,vanHateren_92}, but at all orders of correlation; in other words, the optimal code will minimize the mutual information between different neurons at the output. This led to the idea of `independent component analysis' (ICA) and sparse codes. Imagining that neurons in the visual pathway perform linear filtering on the incoming stimuli and optimizing the filters using ICA over natural scenes, two groups recovered the well-known Gabor-like filters found in the primary visual cortex \cite{olshausen+field_96,bell+sejnowski_97}. 

In the presence of nonlinearities and noise, optimizing information is much more difficult; early work in this direction was done by Linsker \cite{linsker_88}. If neurons are noisy, perfect decorrelation is no longer optimal. In this case, some amount of redundancy (in engineered codes in the form of error-correcting bits) can be beneficial \cite{barlow_01}. It is interesting to note that real neuronal spike trains are usually weakly correlated at the pairwise level and redundant \cite{Puchalla}, and at the level of a large population this redundancy can suffice to predict well the behavior of one neuron from the state of the rest of the network \cite{tkacik+al_14}. If noise is small, information optimization can be carried out in a continuous nonlinear system, and the tradeoffs between decorrelation and redundancy reduction can be explored in detail; in this framework, Karklin and Simoncelli managed to derive many properties of the retina simultaneously from the statistical structure of natural scenes \cite{karklin+simoncelli_11}. This optimization, as well as previous work \cite{balasubramanian+al_01}, also incorporated an important constraint on metabolic efficiency (cost of spiking). By departing from analytically tractable cases, numerical exploration has been used to study optimal coding at arbitrary noise levels with high-dimensional stimuli \cite{tkacik+al_10}.  Using a model for an interacting neural population that can be linked to maximum maximum entropy models with an Ising form (Section \ref{maxent}), small networks experienced a transition between redundant coding and decorrelation based on the statistical structure of the inputs and the level of noise in single units. Surprisingly, when noise in single units is high enough, the redundant strategy with information-maximizing connectivity yields a network with strong attractor states to which different stimuli map uniquely, recovering Hopfield-like associative memory \cite{hopfield_82} from an optimization principle. Mapping out a complete `phase diagram' of optimal coding strategies for high-dimensional noisy nonlinear input/output maps remains a challenge.

Finally, we note that many popular models for coding by large populations of neurons are inefficient in the extreme. It is widely assumed that neurons in a single region of the visual cortex, for example, are selective for a very limited number of features (e.g., the velocity of motion, or the orientation of an edge).  If all the neurons in the population respond independently to these features, with each of the $N$ cells preferring a slightly different value, and we assume that what matters is the number of spikes each neuron generates in a relatively large window of time (``rate coding''), then the response of the population as a whole will allow us to determine the value of the feature with a precision $\sim 1/\sqrt{N}$.  But this means that the information conveyed by the population is $\sim \log(N)$, and thus the information per neuron vanishes as $N$ becomes large.  Are we missing something?   One possibility is that the diversity of dynamical responses among neurons with redundant static feature selectivity can be sufficient to create a combinatorial code for stimulus timing, allowing the information to grow linearly with the number of neurons over a much larger dynamic range \cite{osborne+al_08}.  A more radical alternative is suggested by the discovery of ``grid cells,'' neurons that encode an animal's position relative to a lattice \cite{hafting+al_05}.  In this system, it is possible for a population of neurons with different lattice spacing to generate a truly efficient code, with more neurons allowing exponentially more accuracy in the representation of position \cite{fiete+al_08,sreenivasan+fiete_11}.  These ideas point to the need for direct measurement of the information carried by large populations of cells, under reasonably natural conditions. 

\subsection{Biochemical and genetic networks}

Even the simplest of bacteria have genes that code for many hundreds of proteins, and our own genome codes for more than ten thousand proteins.  No single cell needs to make all these proteins at once, and there are many layers of regulation to control the readout or expression of individual genes.  One important class of these mechanisms is the binding of protein molecules, called transcription factors, to specific sequences along the DNA, close to the start of the coding region of a gene.  Transcription factor binding can both repress or activate protein synthesis; in bacteria there are simple geometrical pictures for how this can happen that are probably good approximations to reality, but the situation in higher organisms is less clear.  In all cases, transcription factors act at concentrations in the nanoMolar range, and $1\,{\rm nM} \sim 1\,{\rm molecule/}\mu{\rm m}^3$.  But a bacterium has a volume of only a few cubic microns, which means that the regulation of transcription is controlled by just handfuls of molecules.

The small numbers of molecules involved means that the regulation of gene expression is, inevitably, noisy, and this noise was measured directly in the early 2000s \cite{elowitz+al_02,ozbudak+al_02,blake+al_03}.  How does this noise limit the transmission of information through the system, from the input concentration of transcription factors to the output level of the protein encoded by the target gene? Here information provides a measure of regulatory power, counting the number of reliably distinguishable output states that can be accessed by adjusting the inputs \cite{tkacik+al_08b,tkacik+walczak_11}.  We recall that if the input/output relation and noise in the system are fixed, then we can optimize information transmission by adjusting the distribution of input signals.  This matching condition, which is essentially the same as in Laughlin's discussion of the fly retina, allows the system to get as much regulatory power as possible out of a limited number of molecules.  An important difference from Laughlin's discussion is that noise in the case of transcriptional regulation must be (and is measured to be) strongly dependent on the levels of input and output.

Early events in development of the fruit fly embryo provide a productive testing ground for ideas about the regulation of gene expression.  During the construction of the egg, the mother places the mRNA for a transcription factor at the end  that will become the head; the combination of synthesis, diffusion and degradation of the protein (Bicoid) translated from this mRNA results in a spatially varying concentration along the length of embryo---a gradient in the concentration of a primary morphogen, to use the conventional language.  Bicoid is the input to an interacting network of genes (called ``gap genes'' and ``pair-rule genes'') that end up being expressed in distinct spatial patterns, which will define the final body plan of the larval fly.  Importantly, these patterns are visible within a few hours after the egg is laid, before cells start to move and even before there are membranes marking the boundaries between the cells.  

A first generation of quantitative experiments on this system mapped the relationship between the input transcription factor concentration and the output level of one of the gap genes, making measurements nucleus by nucleus in single embryos, and demonstrating that the results are consistent across embryos \cite{gregor+al_07a,gregor+al_07b}.  If we think of $x$ as the input concentration and $y$ as the output concentration, then these experiments allow us to estimate the behavior of the communication channel, $P(y|x)$.  We can solve numerically for  the distribution of inputs that optimizes information transmission, and compare with experiment.  Because the mean input/output relation has broad, flat segments, many different input distributions give essentially the same mutual information between $x$ and $y$, but all these near--optimal input distributions have the same output distribution, so we make the comparison of the predicted $P_{\rm out}(y)$ with experiment \cite{tkacik+al_08a}.  The agreement, which involves no adjustable parameters, is excellent.  Importantly, we can also estimate directly the mutual information $I(x;y)$ from the data, and find that it is $0.88\pm 0.09$ of the optimum, so that the system really is transmitting nearly the maximal information given the measured noise levels.  This information is about a bit and a half, showing that the regulation of gene expression in this case is more than just an on/off switch.

Thinking of transcriptional regulation in the early embryo as having one input and one output is a drastic oversimplification, but dealing with multiple inputs and multiple outputs is very difficult.  It is thus useful to note that the information of relevance for development is information about the position of the cell.  If we can measure the expression levels of multiple genes simultaneously, we can ask how much information these level provide about position even if we can't also measure all the input signals.  Indeed, the qualitative concept of positional information is very old \cite{wolpert_69}, and it is attractive to try and make this quantitative \cite{dubuis+al_13b,tkacik+al_14c}.  Refinement of classical staining methods make it possible to measure, simultaneously and with high accuracy, the normalized concentrations of four gap genes that provide dominant signals along the central 80\% of the embryo's length \cite{dubuis+al_13a}.  Analysis of these experiments shows that is is possible to ``decode'' the expression levels and estimate the position of a nucleus to an accuracy of better than 1\% of the length of egg, corresponding  to a total of $4.14\pm 0.05$ bits of information about position \cite{dubuis+al_13b}.    This precision is comparable to the spacing between neighboring cells, and to the precision of subsequent developmental events, suggesting that the embryo really does use this information.  

Each of the four gap genes has a complicated profile of concentration vs. position along the embryo, and the noise levels have an even more complicated structure.  But since cells are almost uniformly distributed, matching of the distribution on inputs to the noise characteristics of the system would require that the error in estimating position be almost uniform as well.  This is, in fact, what we see in the data \cite{dubuis+al_13b}; indeed, if we imagine trying to redistribute the cells so as to transmit more information, we could gain less than 2\%.  This seems a striking signature of optimization.

Optimization of information flow should involve more than matching: we should be able to find the architecture and parameters of genetic networks that allow for maximal information transmission.  One ambitious idea is to define a dynamics in the space of networks that is a schematic of the evolutionary process, and then use information as a surrogate for fitness \cite{francois+siggia_10}.   The networks that emerge from such simulations have many qualitative features in common with the networks functioning in real embryos, including the fruit fly.  One can find examples of multiple genes driven by a common maternal morphogen, each one activating itself and repressing other genes in the network, much as for the gap genes.  A conceptually striking feature of this work is that it explicitly constructs paths through the space of possible networks that lead from simpler to more complex body plans, through gene duplication followed by mutation.  Importantly, almost all the steps along the path are either neutral or increase the fitness.  As the authors emphasize, this ``... provides a quantitative demonstration that continuous positive selection can generate complex phenotypes from simple components by incremental evolution, as Darwin proposed.''  

Rather than reconstructing entire evolutionary trajectories for networks, we can try to find locally optimal networks within some broad class of models \cite{genesI,genesII,genesIII}.    Here again we can see features that remind us of real networks, but there are surprises.  When the available concentration of transcription factors is severely limited, the optimal strategy for transmitting information is for a single transcription factor to broadcast to multiple, completely redundant targets, essentially using the redundancy to make multiple measurements of a weak signal.  Observations of redundancy in genetic networks usually is taken as prima facie evidence against optimization and in favor of an interpretation in which network architecture is just an artifact of evolutionary history, but here we have a quantitative demonstration that such qualitative arguments can be wrong.  More subtly, these calculations show that even qualitative features of the networks which maximize information transmission depend on some molecular details of the regulatory mechanisms.  This is slightly worrisome, but also suggests a path for some of the intricacy of complex molecular interactions to influence macroscopic behavior.    A related approach to these same problems is inspired not by embryos but by artificial genetic circuits that have more limited topologies and inputs, allowing for an exhaustive exploration \cite{ziv+al_07}.  In these cases, the optimization of information transmission leads to very broad maxima in the space of network parameters, so that network topology is the dominant consideration.   These efforts at full optimization have been limited to static inputs and outputs;  information transmission in dynamic contexts raises  new questions, which are just starting to be explored \cite{tostevin+wolde_09,ronde+al_10}.  Finally, interesting questions that have been highlighted recently include the link between information transmission and cellular decision making \cite{bowsher+swain_14}, and using information theory to put strict bounds on the reliability of biochemical processes even when many of these processes are unknown or unobserved \cite{lestas+al_10}.

\subsection{More ambitious goals}

Success at gambling, and perhaps at life, plausibly depends on how much information we have.   If we know more about the outcome of a coin flip (or a horse race, or the stock market, or ... ), then we should be able to win our bets more often, and hence make more money. But it is not obvious that the knowledge measured by Shannon's bits quantifies the useful knowledge in this context. Kelly analyzed a simple gambling game and showed that the maximum rate at which winnings can grow is precisely the information that bettors have about the outcome, and his proof is constructive, so we actually know how to achieve this maximum \cite{kelly_56}.  We can think of the choices that organisms make---e.g., to express different levels of proteins---as being bets on the state of the environment, and there have been several efforts to construct scenarios that match the structure of Kelly's argument, so that the growth rate of the organisms becomes equal to the information that they have about environmental conditions \cite{bergstrom+lachmann_05,kussell+leibler_05,rivoire+leibler_11}. More generally, from rate distortion theory we know that achieving a given average growth rate across a wide range of conditions will require some minimum amount of information about the environment \cite{taylor+al_07}.  These connections between growth rates and information provide a path for evolution to select for organisms that gather and represent more information about their environment.

Another interesting connection of information theoretic optimization to biological function is in the problem of search \cite{vergassola+al_07}.  Organisms search for food, for shelter, for mates, and more; it may even be possible to think more generally about computations or decisions as the search for solutions.  If we imagine that we are looking for something in one of $N$ possible places, we can order these by our current estimate of the probability $P_n$ that we will find what we are looking for in the $n^{\rm th}$ place, so that $P_1 > P_2 > \cdots > P_n$.  Then if we look first in the most likely place, and then down the list, the mean number of tries before we find what we are looking for is $\bar n = \sum_{n=1}^N n P_n$.  Given $\bar n$, there is a probability distribution ($P_n \propto e^{-\lambda n}$) that has maximal entropy.  Conversely, if our uncertainty corresponds to a probability distribution with entropy $S$, then the mean search time $\bar n$ has a minimum, independent of all other details.  The only way we can speed up our search is to reduce the entropy of the distribution, or to gain information in the Shannon sense; while there is no guarantee that our search will go as fast as allowed by this information, it cannot go faster.  

The idea of ``infotaxis''  is  to replace the task of finding something with the task of gaining information about its location \cite{vergassola+al_07}.  In the context of a flying insect searching for the source of an odor, where new information arrives only intermittently when the flight path crosses odor plumes, optimizing the gathering of information by properly alternating upwind flight with cross--wind `casting' in fact leads to the source, and the flight paths that are generated by this algorithm are reminiscent of the paths taken by insects in natural environments.\footnote{As an aside, we note that insects seem to ``know'' about the intermittent structure of turbulent flows \cite{mafra-neto+carde_94}.  In particular, moths exposed to smooth ribbon plumes, which provide a continuous signal and should in principle be easy to follow, fly along strongly zig--zag trajectories,  spending most their time casting crosswind, while the same moths exposed to  intermittent or turbulent plumes fly efficiently toward the source of the odor. This is one of the strongest, qualitative results demonstrating that neural processing is tuned to the statistical structure of physical signals that arise in the natural environment.}  For the purposes of this review, perhaps the most important result from the infotaxis idea is the demonstration that a biologically grounded, goal--directed behavior can be replaced with an abstract, information theoretic optimization principle.  This certainly encourages us to think that some of the nitty--gritty of biology should be derivable from more general principles.

In their earliest work, Attneave and Barlow expressed the hope that the search for efficient representations of sensory data would drive discovery of  features and objects that form our perceptual vocabulary, and Attneave even tried simple experiments to show that human subjects use efficient strategies in abstracting simple figures into sketches \cite{barlow_59,barlow_61,attneave_54}.  As an example, if our world consists of a finite set of rigid objects, each seen from many different perspectives, then the most efficient representation of the image data reaching our eyes would consist of a discrete label for object identity and a few parameters to describe the viewing geometry.  In this case, optimization in an information theoretic sense would require our brains to recognize objects, and the labeling of these objects would constitute a `naming' of the meaningful features of the visual world, without any additional instructions.  At its most ambitious, the claim would be that these features acquire meaning because they are maximally informative,  as opposed to having meaning imposed upon them.  

Shannon frequently used English as an example of information theoretic ideas, and even played a ``guessing game'' with human subjects to estimate the entropy of written texts \cite{shannon_51}.  On the other hand, there is a persistent suspicion that there must be more than just statistics involved in the structure and meaning of language, and one of  the foundational papers of modern linguistics is essentially an attack on Shannon's probabilistic models for language \cite{chomsky_56}.  Some of these early arguments are related to the crude Markovian models that Shannon used as illustrations, and some are related to the confusion between learning a probabilistic model and simply counting events,\footnote{Thus, Chomsky famously argued that statistical approaches could not distinguish between sentences that have never occurred but are instantly recognizable as grammatical (Colorless green ideas sleep furiously) and sentences that never occurred because they are forbidden by grammatical rules (Furiously sleep ideas green colorless) \cite{chomsky_56}. As a concrete rejoinder, Pereira constructed a simple Markov model in which successive words belong to clusters, so that $P(W_{\rm n+1}, W_{\rm n}) = \sum_C Q(W_{\rm n+1}|C) Q(W_{\rm n}|C) P(C)$, and learned the model  on a modest body of newspaper texts using just sixteen clusters.  He found that Chomsky's two example sentences  have probabilities that differ by a factor of more than $10^5$ \cite{pereira_00}.   But between the dates of these references, a much more sophisticated view of the learning problem had emerged.}  but questions remain, and Shannon himself eschewed any connection to the problem of meaning.  In what proved to be a more direct attack on this question, Pereira et al \cite{pereira+al_93} developed an algorithm to cluster nouns based on the distribution of verbs that occur in the same sentence; their approach can be thought of as compressing the description of the noun while preserving information about the identity of the verb, and hence is yet another example of efficient representation \cite{tishby+al_99}.  The resulting clusters exhibit a striking degree of semantic (rather than just syntactic) coherence, and it hard to resist saying that this purely information theoretic approach is extracting meaning.  These references are just a small sampling from the roots of a huge literature.  By now, we all interact with probabilistic models of language every day (through internet search engines, automatic translation, ...), so perhaps the larger conceptual question of whether the search for efficient representation captures meaning has become less urgent.  Nonetheless, we suspect that there are many open questions here, including whether we really understand how to write down models that capture the very long--ranged correlations that are present in texts.

The entropy of a probability distribution is defined without reference to anything else, at least in the case of discrete variables, but information is always about something.  Thus it makes no sense to ask about the information content of a signal; one has to ask how much information the signal provides about some quantity of interest.\footnote{We could be interested in the value of the signal itself, in which case the information is equal to the entropy.  But this is seldom the case.  If the signal is the text of a newspaper, for example, we probably are not interested in the text itself, but what it tells us about events in the world.  In particular, the entropy of the text is increased if we introduce random spelling errors, but  the information about events  is decreased.}  Arguably the most successful uses of information theory in thinking about biological systems are in those cases where the question ``information about what?'' has a clear answer.  But this makes us worry that more of the predictive power of our theories resides in this qualitative knowledge of the biologically relevant information than in the more mathematical principle of optimizing information transmission.  Can we do something more general than just making a list of which information is relevant for each particular system?  

Information cannot be useful to an organism unless it carries predictive power:  if we use information to guide our actions, it takes time to act, and hence  any measure of the success of our actions will be in relation to a future state of the world \cite{bialek+al_07}.  While not all predictive information is equally useful, non--predictive information is useless, and so separating predictive from non--predictive information is an important task.   If we observe a system for time $T$, then the distribution from which trajectories are drawn has an entropy $S(T)$ that we expect to be extensive at large $T$.  But the mutual information between, for example,  observations on the past ($-T < t \leq 0$) and and equally long future ($0 < t \leq T$) is given by $I_{\rm pred}(T) = 2S(T) - S(2T)$.  Importantly, any extensive component of the entropy cancels, and the information that the past provides about the future must be sub--extensive, even if we allow for an unlimited future, so most of the bits we observe must be discarded if we want to isolate the predictive bits.   This sub--extensive entropy is related to our intuitive notions of the complexity of the underlying processes \cite{bialek+al_01a,bialek+al_01b}, and providing an efficient representation of predictive information is connected to problems ranging from signal processing to learning \cite{bialek+al_07,creutzig+al_09}.  Recent work provides preliminary evidence that populations of neurons in the retina indeed provide such an efficient representation, separating predictive information from the non--predictive background at least in simplified visual worlds \cite{palmer+al_13}, but much more remains to be done in this direction.

\section{Outlook}

We can trace the use of information theoretic ideas to describe biological systems back roughly sixty years, almost to the origins of information theory itself.  We have emphasized that these connections between information and the phenomena of life have two very different flavors, one grounded in data and the other aiming for a theory in which the behaviors of real biological systems are derivable from some sort of optimization principle.  Both directions have been productive, and as the volume of quantitative experimental data on biological systems grows, it seems clear that the utility of the approaches to data analysis described in Section \ref{sec3} will grow in parallel.  The search for real theories is more ambitious.

It is important that we have concrete examples of biological systems that are operating near an optimum of information transmission or efficiency of representation, as reviewed in Section \ref{sec5}.  What we don't know is whether these are isolated instances, or examples of a general principle.  It is relatively easy to work out what optimization means in simplified settings, but much harder to address the theoretical issues that arise in more realistic contexts, e.g. with many interacting degrees of freedom and the complexities of fully natural signals.  At the same time, it took decades to get to the point where even the most basic quantities---the actual information carried by biological signals---could be measured reliably.   Now that many of these foundational steps have been solidified, we believe that the coming years will see much more meaningful confrontations between theory and experiment, in a wide range of systems.  

\begin{acknowledgments}
We thank our many collaborators and colleagues, whose names appear in the references, for teaching us so much and for making the exploration of these ideas so pleasurable.  Our work was supported in part by the US National Science Foundation (PHY--1305525 and CCF--0939370), by the Austrian Science Foundation (FWF P25651), by the Human Frontiers Science Program, and by the  Simons and Swartz Foundations.
\end{acknowledgments}

\end{document}